\documentclass[a4paper]{article}

\usepackage{ISCSLP2024}
\usepackage[colorlinks,
            linkcolor=red,       
            anchorcolor=blue,  
            citecolor=green,       
            ]{hyperref}
\usepackage{booktabs}
\usepackage{threeparttable}
\usepackage{multirow}
\usepackage{adjustbox}
\usepackage{bbding, pifont, siunitx}
\usepackage{color}
\usepackage[table]{xcolor}

\title{Prototype and Instance Contrastive Learning for Unsupervised Domain Adaptation in Speaker Verification}

\name{
Wen Huang\textsuperscript{1*}, 
Bing Han\textsuperscript{1*}, 
Zhengyang Chen\textsuperscript{1}, 
Shuai Wang\textsuperscript{2}, 
Yanmin Qian\textsuperscript{1}$^{\dagger}$ 
\thanks{
\textsuperscript{*} Equal contribution.
$^{\dagger}$ Yanmin Qian is the corresponding author.
}
}

\address{
\textsuperscript{1}Auditory Cognition and Computational Acoustics Lab \\
MoE Key Lab of Artificial Intelligence, AI Institute \\
Department of Computer Science and Engineering, Shanghai Jiao Tong University, Shanghai, China \\
\textsuperscript{2}Shenzhen Research Institute of Big Data,  The Chinese University of Hong Kong, Shenzhen, China
}
\email{\{holvan, hanbing97\}@sjtu.edu.cn}

\ninept
\begin{document}

\maketitle
\begin{abstract}
Speaker verification system trained on one domain usually suffers performance degradation when applied to another domain. 
To address this challenge, researchers commonly use feature distribution matching-based methods in unsupervised domain adaptation scenarios where some unlabeled target domain data is available. 
However, these methods often have limited performance improvement and lack generalization in various mismatch situations. 
In this paper, we propose \textbf{P}rototype and \textbf{I}nstance \textbf{C}ontrastive \textbf{L}earning (PICL), a novel method for unsupervised domain adaptation in speaker verification through dual-level contrastive learning. For prototype contrastive learning, we generate pseudo labels via clustering to create dynamically updated prototype representations, aligning instances with their corresponding class or cluster prototypes. For instance contrastive learning, we minimize the distance between different views or augmentations of the same instance, ensuring robust and invariant representations resilient to variations like noise. This dual-level approach provides both high-level and low-level supervision, leading to improved generalization and robustness of the speaker verification model.
Unlike previous studies that only evaluated mismatches in one situation, we have conducted relevant explorations on various datasets and achieved state-of-the-art performance currently, which also proves the generalization of our method.
\end{abstract}

\section{Introduction}
Speaker verification (SV) is the task of identifying a person based on the unique features of their voice. Recently, deep learning-based speaker verification systems have achieved excellent performance in ``in the wild" scenarios~\cite{voxceleb1,voxceleb2}. 
However, despite this success in well-matched conditions, models trained on source domain data often suffer significant performance degradation when applied to mismatched target domains.

To address this issue, researchers have proposed various unsupervised domain adaptation (UDA) methods to improve the generalization of speaker embedding extractors to unseen target domains, primarily focusing on feature distribution matching. For example, Correlation Alignment (CORAL) \cite{sun2016deep,lee2019coral+} aligns the second-order statistics of features between source and target domains. Other approaches, like mutual information minimization (MIM) \cite{kang2022mim} and with- and between-class distribution alignment (WBDA) \cite{hu22b_interspeech}, also reduce feature distribution discrepancies. Additionally, adversarial training-based methods \cite{bhattacharya2019generative, chen2020channel} use neural networks in a min-max game to estimate and eliminate these discrepancies.

While feature distribution alignment methods are conceptually straightforward, they face challenges that limit their performance and generalization. First, in speaker verification, differing speaker identities between source and target domains create an open-set problem. Reducing the distance between distributions is beneficial, but making them identical can be problematic. Rohdin et al. \cite{rohdin2019speaker} found that using adversarial training alone in UDA could even degrade performance and Chen et al. \cite{chen20n_interspeech} introduce partially shared network to alleviate this problem. Furthermore, Li et al. \cite{li2022cdma} transformed the open-set distribution matching problem into a closed-set problem, resulting in performance improvements.
Second, traditional methods primarily rely on target domain distribution data without fully leveraging the potential information it contains. While some advanced methods, like self-supervised learning-based approaches~\cite{chen2021self,pl2,mao2023cluster}, attempt to address this by exploiting potential label information with contrastive loss, they often depend on static labels during adaptation, which can limit performance due to noisy labels and the inherent weaknesses of contrastive loss.

\begin{figure*}[ht]
    \centering
    \includegraphics[width=\textwidth]{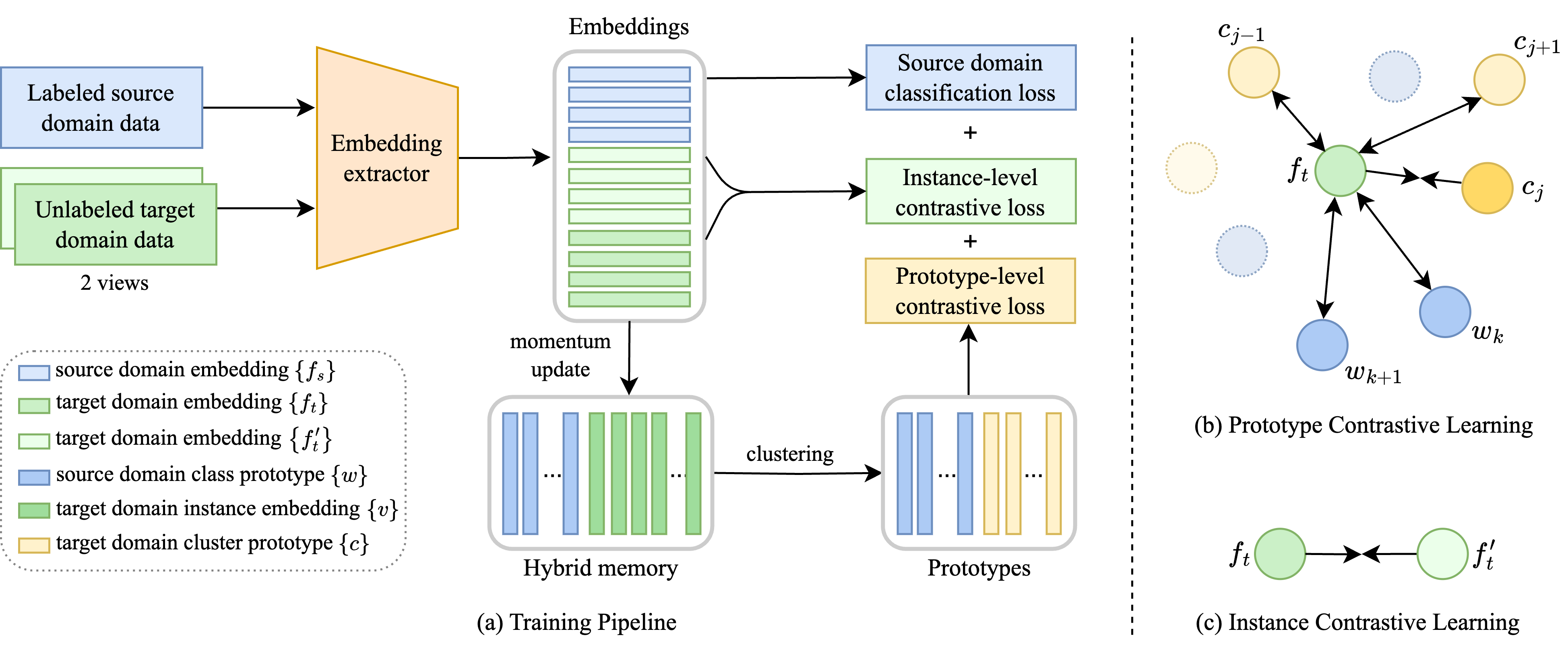}
    \caption{(a) Training pipeline of the proposed Prototype and Instance Contrastive Learning (PICL) for unsupervised domain adaptation. (b) \& (c) Illustration of the dual-level contrastive learning within PICL.}
    \label{fig:framework}
\end{figure*}

In this paper, we propose a \textbf{P}rototype and \textbf{I}nstance \textbf{C}ontrastive \textbf{L}earning (PICL) based unsupervised domain adaptation approach that aims to provide additional supervisory signals for target domain data during the adaptation process. Our method begins with training a speaker embedding extractor on labeled source data, which is then used to generate pseudo-labels for each utterance in the unlabeled target domain via a clustering algorithm. Subsequently, we employ a hybrid memory bank to dynamically update speaker representations, iteratively refining the pseudo-labels for greater accuracy. Based on these pseudo-labels, we perform contrastive learning at both the prototype and instance levels, offering additional supervised signals to enhance performance.
To demonstrate the effectiveness and generalization of our proposed PICL method, we consider a wide range of domain mismatches in real-world scenarios, including language, channel, far- and near-field, and sample rate mismatches. Experimental results show that our method is not only effective but also achieves state-of-the-art performance across all scenarios.

\section{Methods}

To effectively utilize unannotated target domain speaker data and adapt the model accordingly, we implement contrastive learning at two levels: prototype-level and instance-level.

\subsection{Prototype Contrastive Learning}

To address the lack of annotations in the target domain, we leverage pseudo labels generated from clustering algorithms for additional supervision. This is implemented through a two-stage iterative training paradigm. In each training epoch, we first generate pseudo labels for target domain instances via clustering and construct prototypes, then update the model through prototype-based contrastive learning.
Notably, our method differs from previous clustering-based approaches~\cite{mao2023cluster} in two significant ways: (1) the model is trained with supervision from both the source domain and the target domain, and (2) the prototypes are dynamically updated throughout the training process.

Inspired by domain adaptation in object ReID~\cite{ge2020selfpaced}, we construct a hybrid memory in addition to a speaker encoder to store embeddings from both the source and target domains (Fig~\ref{fig:framework}). This hybrid memory dynamically provides and updates prototype embeddings, consisting of source domain class prototypes $\{w_1, …w_{n_s}\}$ and target domain instance embeddings $\{v_1, …v_{n_t}\}$. Here,  $n_s$  denotes the number of classes in the source domain, and  $n_t$  denotes the number of instances in the target domain.

Before training begins, we extract embeddings using the pretrained encoder for both source and target domains, initializing the hybrid memory with source class-average embeddings and target instance embeddings. During training, the embeddings in each mini-batch contribute to updating the hybrid memory in a momentum-based manner, with different update mechanisms for the two parts of the memory. For source domain prototypes, the $k$-th prototype $w_k$ is updated using the average embeddings of class $k$ in the batch:

\begin{equation}
    w_k\gets m^s w_k+(1-m^s)\cdot \frac{1}{|F_k|}\sum_{f^s_i\in F_k}f_i^s
\end{equation}
where $F_k$ is the set of class $k$ embeddings from the source domain in the current mini-batch, and $m^s\in[0,1]$ is the momentum hyperparameter for updating $w_k$.

Since the number of clusters dynamically changes, the target domain cluster prototypes cannot be stored and updated in the same way as the source domain class prototypes. While the hybrid memory caches all target domain instance embeddings $\{v\}$, each embedding $f_i^t$ in the batch can directly update its corresponding instance embedding $v_i$, as follows
\begin{equation}
  v_i\gets m^t v_i+(1-m^t)f_i^t  
\end{equation}
where $m^t \in [0,1]$ is the momentum coefficient used to update the target domain instance embeddings. Consequently, the target domain cluster prototype can be calculated as:
\begin{equation}
    c_j=\frac{1}{|V_j|}\sum_{v_i\in V_j}v_i
\end{equation}
where $V_j$ refers to the set of all updated instance embeddings of the $j$-th cluster. Notably, some clustering algorithms yield outliers, which are treated as separate single-instance classes.

Once the update is complete, we formulate a prototype-level contrastive loss based on the hybrid memory and use this loss to enhance the model’s ability to distinguish between different classes in the source domain and clusters in the target domain. Given an embedding $f$ in the hybrid memory, the prototype-level contrastive loss is defined as follows:

\begin{equation}
\label{eq:lt}
    L_{p}=-\log\frac{\exp(\left\langle f,z^+ \right\rangle/\tau)}{\sum_{k=1}^{n_s}\exp(\left\langle f,w^k \right\rangle/\tau)+
\sum_{k=1}^{n_t^c}\exp(\left\langle f,c^k \right\rangle/\tau)}
\end{equation}
Here $z^+$ represents the positive class prototype corresponding to $f$. If $f$ is a target domain embedding assigned to cluster $k$, then $z^+ = c_k$, the cluster centroid; If $f$ is a source domain embedding belonging to class $k$, then $z^+ = w_k$, the class prototype. $\tau$ is a temperature parameter, $n_s$ and $n_t^c$ denote the total number of source domain classes and target domain clusters, respectively. The inner product $\left\langle \cdot, \cdot \right\rangle$ measures the cosine similarity between two embeddings. 

\subsection{Instance Contrastive Learning}
To complement prototype-level contrastive learning (PCL), we also implement instance-level contrastive learning (ICL). While PCL focuses on aligning instances with their class or cluster prototypes to capture general patterns, ICL ensures the model learns robust and invariant representations for individual instances. Given an input $x$, we generate an additional view $x'$ from the same utterance by applying different augmentations. After obtaining embeddings $f$ and $f'$ for these two views, the instance-level contrastive loss is calculated as follows:
\begin{equation}
    L_i = 1-\left\langle f,f' \right\rangle
\end{equation}
where $\left\langle \cdot, \cdot \right\rangle$ still represents the cosine similarity.
By reducing the distance between different views or augmentations of the same instance, ICL enhances the model’s ability to generalize across various transformations and conditions. 

Furthermore, to ensure consistent classification of the source domain data, we continue using the source classification loss  $L_s$  from the pretrained stage. The final loss is thus defined as $L = L_s + L_p + \lambda L_i$, where $\lambda$ is a hyperparameter used to balance optimization of prototype and instance. 
This dual-level contrastive adaptation improves the model’s overall performance, making it more discriminative and robust for speaker verification tasks.


\section{Experiments and Results}
\subsection{Experimental Settings}
\label{ssec:set}
\noindent\textbf{Datasets.} 
To demonstrate the generalization of our proposed method, we adopt four cross-dataset domain adaptation settings to cover various mismatched situations: (1) VoxCeleb2~\cite{voxceleb2} as the source domain and CN-Celeb~\cite{fan2020cn} as the target domain, focusing on language mismatch; (2) SRE04-10 and SwitchBoard as the source domain and SRE16 as the target domain, focusing on channel and language mismatch; (3) VoxCeleb2 as the source domain and VOiCES19~\cite{richey2018voices} as the target domain, focusing on near-field and far-field mismatch; (4) VoxCeleb2 as the source domain and SRE16 as the target domain, focusing on sample rate and language mismatch.

\vspace{1mm}
\noindent\textbf{Data processing.}
For the CN-Celeb training data, short speech segments are combined to form segments longer than 5 seconds, resulting in 53,288 processed segments. We utilize 80-dimensional Fbank features, with a frame length of 25 ms and a hop size of 10 ms, as the model’s input. The data undergoes preprocessing with three types of augmentation: (1) noise addition using MUSAN~\cite{musan}; (2) reverberation addition using RIRs~\cite{rir}; and (3) speed perturbation~\cite{idlab2020}.

\vspace{1mm}
\noindent\textbf{System configuration.}
The encoder used in the experiments is an r-vector-based ResNet34~\cite{rvector} with 32 channels, and the extracted embedding dimension is set to 256. During the training process, a stochastic gradient descent (SGD) optimizer is employed, with the learning rate decreasing exponentially from 1e-3 to 1e-5. The encoder is initially pre-trained on the source domain data using AAM softmax loss~\cite{aam} with a scale of 32 and a margin that increases from 0 to 0.2. This loss is maintained during the adaptation training stage, but with a fixed margin. 
For PICL, we use DBSCAN~\cite{ester1996density} as the clustering algorithm. The momentum coefficients $m^s$ and $m^t$ are both set to 0.2, and the temperature parameter $\tau$ is set to 0.05.

\vspace{1mm}
\noindent\textbf{Evaluation metrics.}
Cosine similarity is used as the scoring method for CN-celeb and VOiCES19 datasets. For SRE16, the Adapt-PLDA scoring method from the Kaldi toolkit is used, with the PLDA model trained on SRE16-major data. System performance is evaluated using the equal error rate (EER) and the minimum detection cost function (minDCF) with $C_{FR} = 10$, $C_{FA} = 1$, and $P_{target} = 0.01$.

\subsection{UDA Results from Voxceleb to CN-Celeb}
Language mismatch is a very common challenge when applying speaker verification. In this section, we first give a comparison of our proposed PICL with other popular existing methods in Table~\ref{tab:res_cnceleb}. which demonstrates the simplicity and effectiveness of our method. Whether based on distribution-align~\cite{sun2016deep, yi22_interspeech}, self-supervised training~\cite{chen2021self, pl2}, or pseudo labeling methods, we have achieved the SOTA (state-of-the-art) results under the same or larger backbone (Pseudo labeling~\cite{mao2023cluster} and EDITnet~\cite{yi22_interspeech} both adopt larger networks), with a significant improvement. 
Moreover, compared to previous methods~\cite{mao2023cluster} that used static pseudo labels, our dynamic pseudo labels are iteratively updated during the training process. The better results fully demonstrate the necessity of these dynamic updates.

\vspace{-2mm}
\begin{table}[ht]
  \caption{\textbf{Language Mismatch}: Performance comparison of UDA from Voxceleb to CN-Celeb. Results are given with EER(\%) and minDCF. ECAPA-L denotes ECAPA-TDNN with 1024 channel (14.65M Parameters) and ResNet34 only has 6.63M parameters.}
  \label{tab:res_cnceleb}
  \centering
  \begin{tabular}{lcrr}
    \toprule
    \textbf{Method} & \textbf{Backbone} &  \textbf{EER} & \textbf{minDCF} \\
    \midrule
    CORAL~\cite{sun2016deep} & ResNet34 & 10.45 & - \\
    EDITnet~\cite{yi22_interspeech} & SE-ResNet34 & 9.60 & - \\
    SSDA~\cite{chen2021self} & ResNet34 & 10.20 & - \\
    CLISDA~\cite{pl2} & ResNet34 & 9.23 & - \\
    Pseudo Labeling~\cite{mao2023cluster} & ECAPA-L& 8.10 & - \\
    \midrule
    PICL & ResNet34 & \textbf{7.71} & \textbf{0.390} \\
    \bottomrule
  \end{tabular}
\end{table}

\vspace{-2mm}
\subsection{Ablation Study}
Next, we present an ablation study to explore the impact of the hyperparameters $\lambda$ and $m$, with the results listed in Table~\ref{tab:ablation}. Experiments are conducted on CN-Celeb. First, we removed the instance-level loss ($\lambda$=0) to examine the effect of dynamically updated hyperparameter $m$ in the memory bank for prototype-level optimization. According to the results, the dynamically updated weights $m$ do not have a significant impact on EER, but achieve the best minDCF at $m=0.5$. Additionally, we explored the impact of instance-level contrastive learning loss in our proposed PICL by adjusting the weight of $\lambda$. The results indicate that with the inclusion of instance-level optimization, there is a significant improvement in performance, which also confirms the necessity of optimizing the model at dual levels simultaneously in our PICL approach.

\vspace{-2mm}
\begin{table}[ht]
\centering
\caption{\textbf{Ablation study of hyperparameters $\lambda$ and $m$.} Results are given with EER(\%) and minDCF.}
\label{tab:ablation}
\begin{tabular}{ccrr}
    \toprule 
      $\textbf{m}$ & $\lambda$ & \textbf{EER(\%)} & \textbf{minDCF} \\ \midrule
      \cellcolor[HTML]{EFEFEF} 0.2 & 0 & 8.39 & 0.415 \\
      \cellcolor[HTML]{EFEFEF} 0.5 & 0 & 8.49 & 0.390 \\
      \cellcolor[HTML]{EFEFEF} 0.8 & 0 & 8.34 & 0.416 \\
      0.5 & \cellcolor[HTML]{EFEFEF} 1 & 8.22 & 0.416 \\
      0.5 & \cellcolor[HTML]{EFEFEF} 5 & \textbf{7.71} & \textbf{0.390} \\
      0.5 & \cellcolor[HTML]{EFEFEF} 10 & 7.88 & 0.407 \\
    \bottomrule
\end{tabular}
\end{table}

\vspace{-4mm}
\subsection{Results of Other Domain Mismatch}
To further demonstrate the generalization and effectiveness of our proposed PICL, we explored mismatch situations including language, channel, far-field, and sampling rate, and provided performance comparisons with other methods.

\subsubsection{Channel and language mismatch: SRE16}
NIST SRE16 is a dataset widely used to evaluate the robustness of models in channel and language mismatch situations. We evaluated our proposed PICL and presented the results in Table~\ref{tab:res_sre}. In SRE16, previous research has often focused on studying how to align the feature distribution of data from different domains, in order to alleviate the performance degradation caused by domain mismatch. Among them, GAN (generative adversarial networks) and their variants are the most commonly used strategies~\cite{bhattacharya2019generative, rohdin2019speaker, xia2019cross, kataria2021deep}. Our proposed PICL is the first to apply a domain adaptation method based on dynamic pseudo labels to the SRE16 dataset and have achieved the SOTA results so far, demonstrating the effectiveness of our method in channel and language mismatch.

\begin{table}[h]
  \caption{\textbf{Channel and language mismatch}: Performance comparison of UDA on SRE16. Results are given with EER(\%).}
  \label{tab:res_sre}
  \centering
  \begin{adjustbox}{width=.45\textwidth,center}
  \begin{tabular}{lrrrcccc}
    \toprule
    \textbf{Method} & \textbf{Pooled} & \textbf{Tagalog} & \textbf{Cantonese} \\
    \midrule
    Kaldi-ivector \cite{kaldi} & 13.08 & 17.87 & 8.23 \\
    Kaldi-tdnn \cite{kaldi} & 8.66 & 12.63 & 4.69 \\
    CoRAL \cite{sun2016deep} & 9.77 & 13.92 & 5.81 \\
    DANSE \cite{bhattacharya2019adapting} & 13.29 & 17.87 & 8.84 \\
    LSGAN \cite{bhattacharya2019generative} & 11.74 & 15.63 & 7.90 \\
    FuseGAN \cite{bhattacharya2019generative} & 10.88 & 14.94 & 6.93 \\
    ADDA \cite{xia2019cross} & 7.50 & -& - \\
    DF-CycleGAN \cite{kataria2021deep} & - & 13.79 & 6.33 \\
    Mul-MMD \cite{lin2020multi} & 8.28 & - & - \\
    WGAN \cite{rohdin2019speaker} & 9.59 & 13.7 & 5.59 \\
    PSN-ADV \cite{chen20n_interspeech} & 8.98 & 12.9 & 5.18 \\
    VDANN \cite{tu19_interspeech} & 8.21 & - & - \\
    MMD-VDANN \cite{tu2020information} & 7.87 & - & - \\
    WBDA \cite{hu22b_interspeech} & 7.16 & 10.28 & 4.11 \\
    SKD+DABN+DAIN \cite{hu2022domain} & 6.95 & 10.13 & 3.64 \\
    \midrule
    PICL& \textbf{5.40} & \textbf{7.78} & \textbf{3.01} \\
    \bottomrule
  \end{tabular}
  \end{adjustbox}
\end{table}

\subsubsection{Near- and Far-field mismatch: VOiCES19}
Near- and far-field mismatch is also a very common situation when building a SV system. Following the setup in~\cite{yi20b_interspeech}, we evaluate the UDA performance of our proposed PICL on the VOiCE19-eval dataset under near and far- field mismatch. It should be pointed out that on this dataset, instance-level loss will lead to performance degradation, so the results presented in Table~\ref{tab:res_voices} only adopt the loss of prototype-level. This degradation may occur because instance-level loss, which focuses on preserving fine-grained details, could amplify noise and other variations in far-field conditions, leading to less robust representations. By comparing with previous methods, our PICL can achieve current SOTA results without using any speaker labels in target domain, even better than 
systems which adapted with labels~\cite{sang2021deaan, yi2022disentangled}. 
\begin{table}[ht]
  \caption{\textbf{Near- and Far-field mismatch}: Performance comparison of UDA on VOiCES19-Eval dataset. $N$ and $Y$ represent whether speaker labels in the target domain are needed during adaptation respectively.}
  \label{tab:res_voices}
  \centering
  \begin{tabular}{lcrr}
    \toprule
    \textbf{Method} & \textbf{Labeled} & \textbf{EER} & \textbf{minDCF} \\
    \midrule
    DSN+MINE \cite{yi20b_interspeech} & N & 6.51 & 0.631 \\
    ADSAN+MINE \cite{yi20b_interspeech} & N & 6.79 & 0.599 \\
    CDMA-align \cite{li2022cdma} & N & 5.75 & 0.475 \\
    CDMA \cite{li2022cdma} & N & 5.62 & 0.468 \\
    DEANN \cite{sang2021deaan} & Y & 5.21 & 0.394 \\
    InfoMax–DSAN \cite{yi2022disentangled} & Y & 5.69 & 0.413 \\
    \midrule
    PICL & N & 5.12 & 0.381 \\
    \bottomrule
  \end{tabular}
\end{table}

\subsubsection{Sample rate mismatch: From Voxceleb to SRE16}
Sample rate mismatch is a common issue in speech-related tasks and can cause significant performance degradation~\cite{zhang2023toward}, but it has not been given sufficient attention in the SV task. Therefore, by using Voxceleb (16kHz) as the source domain and SRE16 (8kHz) as the target domain, we explore whether our proposed method can effectively alleviate the negative impact of sampling rate mismatch. 
\begin{table}[ht]
    \centering
    \caption{\textbf{Sample rate mismatch}: Performance comparison of UDA using PICL from Voxceleb(16 kHz) to SRE16(8 kHz). Results are given with EER(\%).}
    \label{tab:res_sr}
    \begin{tabular}{r@{\,}lcrrr}
    \toprule
    \textbf{Src} & $\rightarrow$ \textbf{Tgt} & \textbf{Adapt.} & \textbf{Pooled}  & \textbf{Tagalog}  & \textbf{Cantonese} \\
    \midrule
    16k & $\rightarrow$ 8k &  & 12.88 & 17.14 & 7.80 \\
    16k & $\rightarrow$ 16k & \ding{55} & 7.45 & 11.15 & 3.81 \\
    8k & $\rightarrow$ 8k &  & 6.41 & 9.91 & 2.97 \\
    \midrule
    16k & $\rightarrow$ 8k &  & 8.69 & 11.91 & 5.34 \\
    16k & $\rightarrow$ 16k & \checkmark  & 7.49 & 10.89 & 4.11 \\
    8k & $\rightarrow$ 8k &  & 5.63 & 8.59 & 2.75 \\
    \bottomrule
    \end{tabular}
\end{table}

\vspace{-2mm}
According to the results listed in Table~\ref{tab:res_sr}, compared to directly evaluate on 8kHz data using the model pretrained with 16kHz data, resampling the training and testing data to the same frequency(SRE16 to 16kHz or Voxceleb to 8kHz) will result in a significant improvement. Building on this, we incorporated our proposed PICL, which further improved performance and demonstrated the effectiveness of PICL in addressing sampling rate mismatch.

\section{Conclusions}
In this paper, we propose Prototype and Instance Contrastive Learning (PICL) to address domain mismatch in speaker verification. 
Prototype contrastive learning generates pseudo labels via clustering, creating dynamic prototype representations that align instances with their class or cluster. Instance contrastive learning minimizes the distance between different views of the same instance, ensuring robust representations against variations. This dual-level approach improves both high-level and detailed learning, resulting in a more robust speaker verification model.
We have considered various mismatches in real-word scenarios, including language, channel, far- and near-field, sample rate mismatch. The experimental results show that our PICL can achieve the SOTA performance across all scenarios and also demonstrate its generalization capability.

\section{Acknowledgements}
This work was supported in part by China NSFC projects under Grants 62122050 and 62071288, in part by Shanghai Municipal Science and Technology Commission Project under Grant 2021SHZDZX0102.


%


\bibliographystyle{IEEEtran}



\end{document}